\documentclass[conference]{IEEEtran}
\IEEEoverridecommandlockouts
\usepackage{amsmath}
\usepackage{graphicx}
\usepackage{array}
\usepackage[table]{xcolor}
\usepackage{booktabs}
\usepackage{relsize}
\usepackage{soul}

\usepackage{cite}
\usepackage{amsmath,amssymb,amsfonts}
\usepackage{algorithm}
\usepackage{algpseudocode}
\usepackage{graphicx}

\usepackage{textcomp}
\usepackage[colorlinks=true, linkcolor=blue, citecolor=blue, urlcolor=blue]{hyperref}

\usepackage{bm}

\usepackage[caption=false,font=footnotesize]{subfig} % subfigures w/ IEEE captions
\usepackage{stfloats}     % allow two-column floats at bottom (figure* [!b])

\usepackage{xcolor}
\def\BibTeX{{\rm B\kern-.05em{\sc i\kern-.025em b}\kern-.08em
    T\kern-.1667em\lower.7ex\hbox{E}\kern-.125emX}}

\begin{document}

\title{Configuration Tuning for ISAC: Cost-Efficient Adaptation via RACE-CMA} %Configuration tuning for ISAC Adaptive Threshold Optimization for Smart Sensing Feedback in ISAC Systems}

\author{
\IEEEauthorblockN{
Ashkan Jafari Fesharaki\IEEEauthorrefmark{1},
Yasser Mestrah\IEEEauthorrefmark{2},
Ibrahim Hemadeh\IEEEauthorrefmark{2},
Yi Ma\IEEEauthorrefmark{1}, 
Mohammad Heggo \IEEEauthorrefmark{2},\\
Arman Shojaeifard\IEEEauthorrefmark{2},
Ahmet Serdar Tan\IEEEauthorrefmark{2},
Rahim Tafazolli\IEEEauthorrefmark{1},
and Alain Mourad\IEEEauthorrefmark{2}
}
\IEEEauthorblockA{\IEEEauthorrefmark{1}6GIC, Institute for Communication Systems, University of Surrey, Guildford, UK, GU2 7XH}
\IEEEauthorblockA{\IEEEauthorrefmark{2} 
InterDigital, London, UK, EC2A 3QR}
}

\maketitle

\begin{abstract}

This paper studies a feedback-driven configuration-tuning framework for adaptive sensing feedback in Integrated Sensing and Communication (ISAC) systems. We propose a framework in which the User Equipment (UE) adapts sensing parameters under dynamic conditions while satisfying network-defined constraints. The problem is formulated as a stochastic constrained optimization problem, to improve sensing reliability and latency. We consider a bistatic ISAC sensing-feedback setup and instantiate the framework via threshold optimization as a representative case study, enabling benchmarking against baseline methods. To ensure efficiency under UE computational limits, we propose Ranking-Aware, Constrained, and Efficient CMA-ES (RACE-CMA), which integrates two-stage racing, common random numbers, noise-aware ranking, and feasible constraint handling. Results show that the proposed approach improves sensing reliability by about 35\% while reducing computational cost by about 25\%, yielding roughly a twofold gain in performance–cost efficiency. This highlights that UE-side configuration tuning is a promising mechanism for enhancing closed-loop ISAC performance under practical system constraints.
\end{abstract}

\begin{IEEEkeywords}
ISAC, sensing, closed-loop feedback, configuration tuning, CMA-ES, optimization
\end{IEEEkeywords}

\section{Introduction}
Sixth-generation (6G) wireless networks are envisioned to integrate sensing, communication, and control into a unified infrastructure~\cite{zhang2021perceptive, saad2020vision, HanMulti2025}. 
Integrated Sensing and Communication (ISAC) is a key enabler of this vision, transforming the radio access network into a distributed sensory system capable of perceiving and interacting with its surroundings~\cite{wei2022toward, Li2024TowardSeamless}. By jointly exploiting communication waveforms for perception, ISAC enables new capabilities in localization, mobility management, and context-aware networking~\cite{wild2021joint, detection2024Ash}, as reflected in ongoing 3GPP activities toward sensing-enabled networks~\cite{3gppTR22870}. 

In current 5G and pre-6G systems, User Equipment (UE) behavior is governed by network-defined Radio Resource Control (RRC) configurations, where parameters such as thresholds and timers are fixed for worst-case conditions. This static design is simple but inefficient for ISAC sensing in dynamic environments, leading to redundant reports, delayed feedback, and frequent RRC reconfigurations \cite{3gpp_ts_38331}. Because sensing performance depends on interference, target mobility, and channel variation, fixed settings cannot ensure reliability or responsiveness \cite{Ahmadipour2025-mm}. 

By dynamically refining reporting thresholds, timers, resource block (RB) density, beam-sweep cadence, and sensing periodicity, the UE can maintain stable sensing Key Performance Indicators (KPIs) without repeated RRC Reconfiguration. This approach reduces signaling overhead, improves sensing accuracy and responsiveness, and optimizes resource usage while staying fully compliant with operator policy. In this work, configuration tuning is the overarching mechanism, and we apply it atop the Smart Sensing Feedback (SSF) mechanism \cite{Fesharaki2025-rs} as a concrete use case. In brief, SSF defines how the UE reports target-related information (e.g., detected, lost, null) to the transmitting entity. The tunable configuration vector (e.g., thresholds, timers, sensing periodicities, and beam-sweep cadence) is adjusted within network-defined bounds to maintain sensing KPIs under dynamic conditions. For example, thresholds map power measurements (e.g., Reflected Echo Strength Indicator (RESI), Reference Signal Received Power (RSRP)) to UE actions (e.g., reporting, power scaling, reconfiguration). If set too low, they trigger unnecessary RRC updates; if too high, they cause missed detections and increased latency. As the closed-loop ISAC process is stochastic and non-differentiable, optimizing these parameters requires robust, sample-efficient methods under environmental uncertainty.

Prior ISAC studies have mainly focused on architectures, waveform/beamforming design, and sensing coverage, including recent bistatic and multi-static approaches. While these studies advance physical-layer sensing, they do not address how UE-side sensing parameters should be adapted online in a closed-loop feedback setting \cite{isac_study_survey, zhang2021perceptive}. This creates a gap between sensing capability and practical adaptation. At the same time, optimization-based adaptation has been widely explored in other domains using methods such as CMA-ES \cite{Hanz2023CMA}, SPSA \cite{spall1992multivariate}, and interior-point algorithms \cite{DEMARCHI20231247}, with further developments for noisy settings\cite{CMA_noise}. However, these techniques have not been systematically tailored to configuration tuning in closed-loop ISAC, where evaluations are stochastic, expensive, and constrained by network guardrails.

In this paper, we address this gap by proposing a UE-side sensing configuration tuning in closed-loop ISAC framework and formulating it as a stochastic constrained optimization problem. Our objective is to improve sensing reliability and responsiveness under dynamic conditions while respecting network-defined constraints. We consider a bistatic ISAC sensing-feedback setup and use threshold optimization as a representative case study. To solve this problem efficiently due to UE computational cost limits, we propose a noise-robust and cost-efficient optimizer, termed Ranking-Aware, Constrained, and Efficient CMA-ES (RACE-CMA). The main contributions of this paper are as follows:
\begin{itemize}
\item We develop a feedback-driven configuration-tuning framework for closed-loop ISAC, in which sensing-related parameters are adapted within network-defined bounds to improve specific KPIs.
\item We instantiate this formulation in a bistatic ISAC sensing-feedback setting using threshold optimization and adapt existing methods to this framework, providing a benchmark for evaluating the proposed configuration-tuning framework.
\item We propose RACE-CMA, which enhances CMA-ES with two-stage racing, common random numbers, noise-aware ranking, and feasible-by-construction constraint handling, achieving improved performance–cost trade-offs under stochastic dynamics.
\end{itemize}
\section{System and Threshold-Learning Formulation}
\subsection{System and Signal Model}
\begin{figure}
    \centering
    \includegraphics[width=0.45\textwidth]{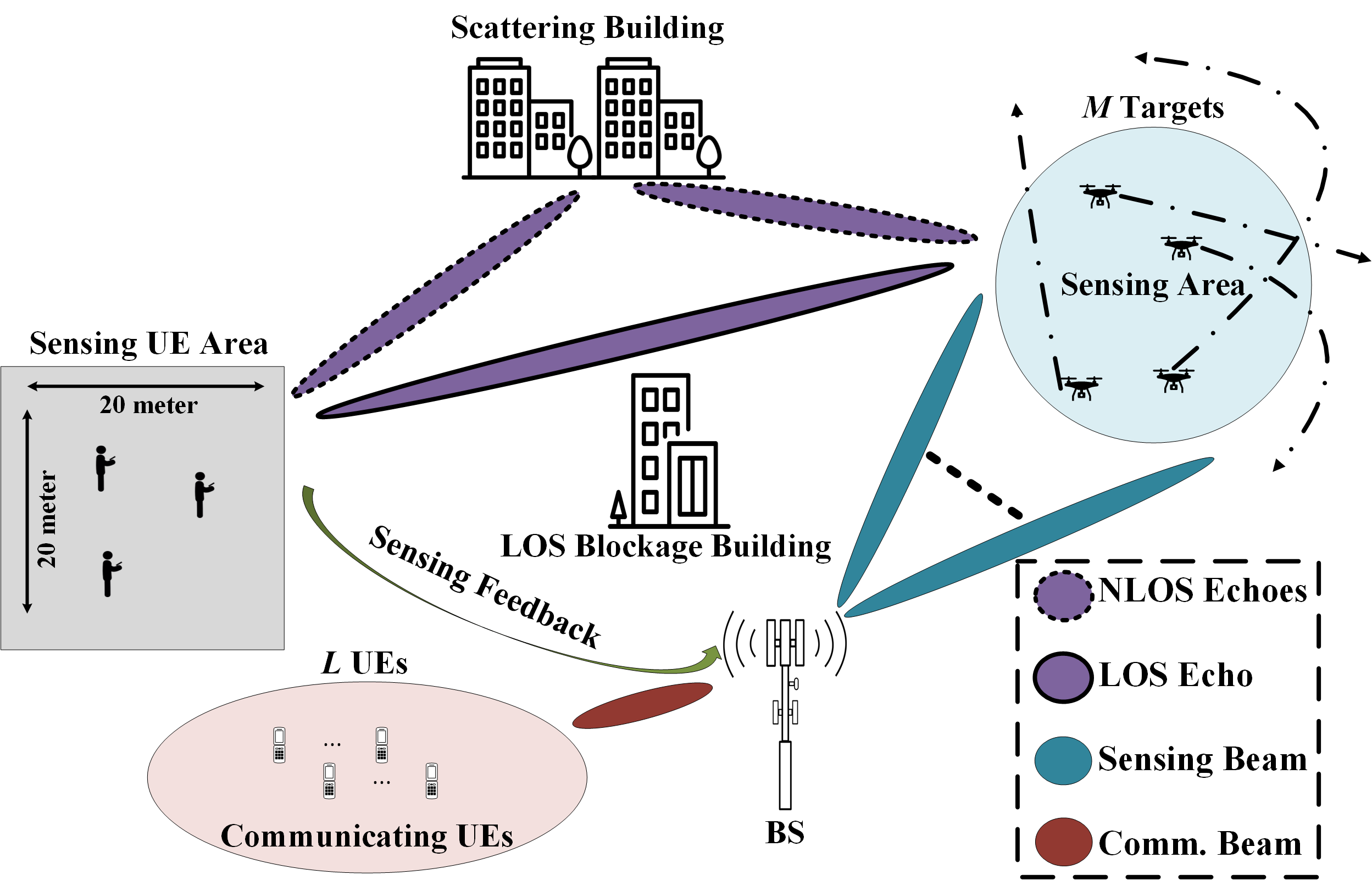}
    \caption{BS serves multiple UEs, performing bistatic sensing in region $\mathcal{D}_S$}
    \label{fig:environment}
\end{figure}

We study a bistatic ISAC setup with one BS transmitter and one sensing UE receiver as illustrated conceptually in Fig.~\ref{fig:environment}. The BS sends OFDM waveforms for joint communication and sensing while serving other UEs. The sensing UE measures echoes from moving targets within the BS sensing region, denoted by $\mathcal{D}_S$. Echoes may arrive through both line-of-sight (LoS) and non-LoS (NLoS) paths created by surrounding scatterers (e.g., buildings, walls), enabling sensing even when the direct path is blocked. The BS employs a narrowband transmit beamforming vector $\mathbf{f}\!\in\!\mathbb{C}^{N_{\text{BS}}\!\times\!1}$, while the UE applies a combining vector $\mathbf{w}\!\in\!\mathbb{C}^{N_{\text{UE}}\!\times\!1}$ with unit norms. Let $X[k,m]$ denote the known pilot on subcarrier $k$ and OFDM symbol $m$, with $k\!\in\!\{0,\dots,N_{\text{sc}}\!-\!1\}$ and $m\!\in\!\{0,\dots,N_{\text{sym}}\!-\!1\}$.  
The received signal at the UE can be expressed as $
Y[k,m] = 
\sqrt{p} \;\!\sum_{t=1}^{N_T}\!
\!\left(\mathbf{w}^{\mathrm{H}}\mathbf{h}^{(\mathrm{ret})}_{k,m,t}\right) 
\!\left(\mathbf{h}^{(\mathrm{fwd})}_{k,t}{}^{\mathrm{H}}\mathbf{f}\right)
X[k,m] + I[k,m] + Z[k,m],$ where $p$ is the sensing power, $Z[k,m]\!\sim\!\mathcal{CN}(0,\sigma^2)$ models thermal noise, and $I[k,m]$ collects clutter and interference.  
The terms $\mathbf{h}^{(\mathrm{fwd})}_{k,t}$ and $\mathbf{h}^{(\mathrm{ret})}_{k,m,t}$ represent the forward (BS$\!\rightarrow$target) and return (target$\!\rightarrow$UE) channels for target $t$, respectively:
\begin{subequations}\label{eq:bistatic-channels}
\begin{align}
& \mathbf{h}^{\mathrm{(fwd)}}_{k,t}
= \sqrt{\Gamma^{\mathrm{BS}}_{t}}\,
  e^{-j2\pi k\Delta f\,\tau_{bt}}\,
  \mathbf{u}_{\mathrm{BS}}\!\big(\varphi_t\big), \\ \notag \\
&\mathbf{h}^{\mathrm{(ret)}}_{k,m,t}
= \sqrt{\Gamma^{\mathrm{UE}}_{t}}\;
   e^{-j2\pi k\Delta f\,\tau_{tu}}\;
   e^{+j2\pi \nu_{t}\,mT_{\mathrm{sym}}}\;
   \mathbf{u}_{\mathrm{UE}}\!\big(\theta_{t}\big) \notag \\
&+ \sum_{\ell=1}^{L_{\mathrm{NLoS}}} \sqrt{\Gamma^{\mathrm{UE}}_{t, \ell}}\;
   e^{-j2\pi k\Delta f\,\tau_{tu, \ell}}\;
   e^{+j2\pi \nu_{t, \ell}\,mT_{\mathrm{sym}}}\;
   \mathbf{u}_{\mathrm{UE}}\!\big(\theta_{t, \ell}\big),
\end{align}
\end{subequations}

where $\tau_{bt}$ and $\tau_{tu}$ denote the propagation delays, $(\varphi_t,\theta_t)$ are the angles of departure and arrival, $\nu_{(\cdot)}$ is Doppler shifts across OFDM symbols, $\mathbf{u}_{\mathrm{BS/UE}}(\cdot)$ is the BS/UE array steering vector and $(\tau_{tu,\ell},\nu_{t,\ell},\theta_{t,\ell})$ are effective NLoS components\footnote{If only one effective NLoS path is retained, set $L_{\mathrm{NLoS}}=1$; if none, drop the sum.}. Additionally, $\Gamma^{\mathrm{BS}}_{t}(\cdot)$, $\Gamma^{\mathrm{UE}}_{t}(\cdot)$ and $\Gamma^{\mathrm{UE}}_{t, \ell}(\cdot)$ correspond to path loss and effective scattering gain.

Stacking all resource elements, the vector observation can be written as $
\mathbf{y} = \sqrt{p}\sum_{t=1}^{N_T}\mathbf{D}(\tau_t,\nu_t;\mathbf{f},\mathbf{w})\mathbf{x} + \mathbf{i} + \mathbf{z}, $ where $\mathbf{D}(\tau_t,\nu_t;\mathbf{f},\mathbf{w})$ encodes the delay–Doppler response of the target.  
A two-dimensional matched filter extracts the reflected component as in $S(\tau,\nu) = \frac{\mathbf{g}(\tau,\nu)^{\!H}\mathbf{y}}{N_{\text{sc}}N_{\text{sym}}},$ with $\mathbf{g}(\tau,\nu)$ the delay–Doppler atom (filter).  
The peak correlation in a search window $\mathcal{W}$ gives the estimated delay–Doppler pair $(\hat{\tau},\hat{\nu})$ and its normalized magnitude is called the RESI, which acts as a scalar statistic summarizing the echo strength and serves as the input to the sensing-feedback controller, resulting in a compact control input rather than a full estimator output:
\begin{equation}
\text{RESI} = \frac{|S(\hat{\tau},\hat{\nu})|}{\hat{\sigma}_Z}, \qquad
\hat{\sigma}_Z = \sqrt{\tfrac{1}{|\mathcal{N}_0|}\sum_{(k,m)\in\mathcal{N}_0}|Z[k,m]|^2},
\end{equation}

\noindent \textbf{Practical assumptions and scope.}
In this study, we assume access to BS pilots, beam context, and network-assisted synchronization, with residual impairments absorbed into the effective disturbance and reflected in the RESI statistic. This aligns with our focus on configuration tuning, while detailed PHY modeling is left for future work.

\subsection{Problem Formulation}

We define the stochastic optimization problem as:
\begin{align}
    \min_{\mathbf{P} \in \mathcal{C}} \quad & \mathbb{E}_{\xi}\left[ \mathcal{J}(\mathbf{P}, \xi) \right]  \\
    \text{subject to} \quad & \mathbf{P} \in \mathcal{C}
    \label{eq:general}
\end{align}
where $\xi$ encapsulates random channel and traffic dynamics (e.g., mobility, blockage, interference), $\mathbf{P}$ will be the configuration parameters (e.g., thresholds, timers, sweep periodicity, and reference-signal (RS) density) configured using AS/non-AS (NAS) configuration, and $\mathcal{J}(\cdot)$ is a multi-objective cost function composed of different objectives. Moreover, the feasible set $\mathcal{C}$ is defined by the network guardrails, such as minimum/maximum threshold values, timer bounds, or resolution granularity. 

In the following, without loss of generality, we will focus on formulating the optimization problem using a subset of $\mathbf{P}$, which is the decision thresholds ($\mathbf{T}$). For this purpose, we use the adaptive sensing feedback mechanism~\cite{Fesharaki2025-rs}, which defines a generalized closed-loop framework through which the sensing UE interacts with the network via a set of discrete sensing states. This framework acts as a multi-hypothesis detector over $K$ hypotheses $\{\mathcal{H}_0,\dots,\mathcal{H}_{K-1}\}$, each defined by: (i) decision thresholds $\{T_0,\dots,T_{K-2}\}$, (ii) activation conditions $\{\mathcal{C}_i\}$ (e.g., power, doppler, interference), and (iii) actions $\{\mathcal{A}_i\}$ determining the UE’s reporting or sensing mode.  
This structure allows adapting to mobility and channel variations while staying aligned with network procedures. As an instance, when the measurement metric is the RESI, $x_t$ at frame $t$, the UE maps $x_t$ to one of four states via $\mathcal{H}_i : T_{i-1} < x_t \le T_i$ with $i=0,1,2,3 \,$, where $T_{-1}=-\infty$, $T_3=\infty$, and tunable thresholds $\mathbf{T}=[T_1,T_2,T_3]$. Each state $\mathcal{H}_i$ triggers a specific feedback action (e.g., power scaling, beam update, sensing periodicity), shaping the closed-loop ISAC behavior. Based on this, we define the following objective functions to be integrated in the multi-objective function $\mathcal{J}(\cdot)$ in Eq. \ref{eq:general} :
\begin{itemize}
\item Detection reliability:
\begin{equation}
J_{\mathrm{det}}(\mathbf{T}) = \frac{\sum_{t=1}^{T_{\text{S}}}  [tg(t)\in Beam(\mathbf{T},t)] \wedge [x_t > T_1]
}{\sum_{t=1}^{T_{\text{S}}} [tg(t) \in \mathcal{D}_{\text{S}}]},
\label{eq:p-det}
\end{equation}
Here, $J_{\mathrm{det}}$ denotes a coverage-aware detection reliability metric, rather than a conventional physical-layer probability of detection under a fixed false-alarm constraint.
\item Sensing latency and report age: 
\begin{equation}
J_{\mathrm{lat}}(\mathbf{T})
=  
\begin{cases}
T_{\text{S}}, \qquad{\sum_{t=1}^{T_{\text{S}}} \scriptstyle{[x_t > T_1]} = 0} \\
\sum_{i=1}^{T_{\text{S}}}  t_{[\mathcal{H}_3 \rightarrow \mathcal{H}_j]}^{i} \; - \; t_{[\mathcal{H}_j \rightarrow \mathcal{H}_3] }^{i},  &\scriptstyle \text{Otherwise}
\end{cases}
\end{equation}
\item Power and processing overhead:
    \begin{equation}
\ J_{\mathrm{pow}}(\mathbf{T})
= \frac{1}{T_S}\sum_{t=1}^{T_\text{S}}  \eta(x_t,\mathbf{T}) P_{\text{budget}},
\end{equation}
\end{itemize}

This formulation captures the closed-loop nature of the problem: $\mathbf{T}$ affects sensing decisions, which influence system state (e.g., queue length, feedback rate), ultimately feeding back into cost function evaluations. Moreover, this optimization problem is stochastic, and computationally expensive to deploy in the UE since each evaluation requires a full ISAC simulation. Therefore, efficient and noise-robust optimization methods are required to achieve reliable convergence within practical computational and resource budgets.

\section{Proposed Possible Solutions}
\label{sec:baseline}

In this section, we adapt four optimization methods for configuration tuning, encompassing analytic, deterministic, stochastic, and population-based approaches. 

\subsection{Maximum A-Posteriori (MAP) Rule}
The classical MAP rule analytically places each threshold $\eta_i$ where the posterior probabilities of neighbouring hypotheses are equal:
$p(\eta_i|\mathcal{H}_{i-1})P(\mathcal{H}_{i-1}) = p(\eta_i|\mathcal{H}_i)P(\mathcal{H}_i)$ where  $i=1,2,3$. The conditional densities $p(x|H_i)$ and priors $P(H_i)$ can be fitted from one Monte Carlo run and used to compute $\eta_i$ directly.  
MAP provides a deterministic and extremely fast solution requiring only one simulation for density estimation; however, MAP cannot adapt to environmental changes and does not optimize the true stochastic cost $\hat{p}_{\text{det}}(\mathbf{T})$ within the closed feedback loop.

\subsection{Interior-Point Newton (IPN)}
IPN performs deterministic local search using finite-difference gradients and a logarithmic barrier to enforce ordering constraints.  
At iteration $k$, the penalized objective is
$\Phi_{\mu_k}(\mathbf{T}^{(k)}) =
\mathcal{J}(\mathbf{T}^{(k)}) -
\mu_k \big[\log(T_2\!-\!T_1) + \log(T_3\!-\!T_2)\big],$ where $\mu_k>0$ is gradually reduced.  
A Newton step is computed as
\begin{equation}
\begin{aligned}
\nabla^{2}\Phi_{\mu_k}(\mathbf{T}^{(k)})\,\mathbf{d}^{(k)}
   = -\nabla\Phi_{\mu_k}&(\mathbf{T}^{(k)}), \\ \mathbf{T}^{(k+1)}
   = \mathbf{T}^{(k)}+\alpha^{(k)}\mathbf{d}^{(k)},
\end{aligned}
\end{equation}
with $\alpha_k$ found by backtracking line search. The gradient and Hessian are approximated by central differences, requiring $(2n+1)$ objective evaluations per iteration, where $n=3$ thresholds. Therefore, IPN's cost will be $C_{\text{IPN}} = (2n+1) C_f.$
IPN converges quadratically when the objective is smooth, but because $J_{\mathrm{det}}(\mathbf{T})$ is noisy, finite-difference estimates are unstable, causing divergence or convergence to local optima.

\subsection{Simultaneous Perturbation Stochastic Approx. (SPSA)}
SPSA estimates all gradient components with only two evaluations per iteration, independent of dimension.  
For iteration $k$, a random perturbation $\Delta^{(k)}\in\{\!-\!1,1\}^3$ is drawn and a small step $c_k$ is chosen.  
The gradient estimate is
\begin{equation}
\hat{\nabla}\mathcal{J}_i(\mathbf{T}^{(k)}) =
\frac{\mathcal{J}(\mathbf{T}^{(k)}+c_k\Delta^{(k)}) -
\mathcal{J}(\mathbf{T}^{(k)}-c_k\Delta^{(k)})}
{2 c_k \Delta^{(k)}_i},
\end{equation}
and the update is $\mathbf{T}^{(k+1)} =
P_\Omega\!\left[\mathbf{T}^{(k)} - a_k\,\hat{\nabla}\mathcal{J}(\mathbf{T}^{(k)})\right],$ where $a_k$ is a diminishing gain and $P_\Omega[\cdot]$ reorders entries to preserve feasibility.  
Each iteration needs only two full evaluations, therefore SPSA's cost will be $C_{\text{SPSA}} = 2 C_f.$
SPSA offers strong noise tolerance through symmetric perturbations, but the convergence rate is slow ($O(1/k)$), and its stochastic gradient path can oscillate when noise variance is high.

\subsection{Covariance Matrix Adaptation Evolution (CMA-ES)}
CMA-ES is a derivative-free, population-based optimizer that adapts a Gaussian search distribution so that exploration aligns with the local curvature of the objective landscape.  
At generation $g$, a population of $\lambda$ offspring is drawn as $\mathbf{T}^{(j)} \sim 
\mathcal{N}\!\big(m^{(g)},(\sigma^{(g)})^{2}C^{(g)}\big)$ where $j=1,\dots,\lambda$. Then it will be mapped into the feasible set and ranked by $\mathcal{J}(\mathbf{T}^{(j)})$.  
Let $x_{i:\lambda}$ denote the $i$-th best offspring, and define normalized steps $y_i=(x_{i:\lambda}-m^{(g)})/\sigma^{(g)}$ with recombination weights $w_i>0$, $\sum_i w_i=1$, and $\mu_{\mathrm{eff}}=(\sum_i w_i)^2/\sum_i w_i^2$.  
The mean update will be $m^{(g+1)}=\sum_{i=1}^{\mu} w_i\,x_{i:\lambda}.$
% \begin{equation}
% m^{(g+1)}=\sum_{i=1}^{\mu} w_i\,x_{i:\lambda}.
% \label{eq:meanupdate}
% \end{equation}
Two evolution paths track recent successful directions.  
The step-size path measures the length of whitened moves,
\begin{equation}
p_\sigma^{(g+1)}=(1-c_\sigma)p_\sigma^{(g)}
+\sqrt{c_\sigma(2-c_\sigma)\mu_{\mathrm{eff}}}\,
C^{-(g)\!/2}\frac{m^{(g+1)}-m^{(g)}}{\sigma^{(g)}},
\label{eq:pq}
\end{equation}
whose expected norm $\chi_n=\mathbb{E}\Vert\mathcal{N}(0,I)\Vert$ controls global scaling:
\begin{equation}
\sigma^{(g+1)}=\sigma^{(g)}\exp\!\left(
\frac{c_\sigma}{d_\sigma}
\Big(\frac{\|p_\sigma^{(g+1)}\|}{\chi_n}-1\Big)\right).
\label{eq:sig}
\end{equation}
The covariance path accumulates raw-space directions,
\begin{equation}
p_c^{(g+1)}=(1-c_c)p_c^{(g)}
+\sqrt{c_c(2-c_c)\mu_{\mathrm{eff}}}\,
\frac{m^{(g+1)}-m^{(g)}}{\sigma^{(g)}},
\label{eq:pc}
\end{equation}
and the covariance matrix adapts as
\begin{equation}
C^{(g+1)}=(1-c_1-c_\mu)C^{(g)}
+c_1\,p_c^{(g+1)}{p_c^{(g+1)}}^\top
+c_\mu \sum_{i=1}^{\mu} w_i\,y_i y_i^\top.
\label{eq:up}
\end{equation}
Intuitively, $C$ expands along consistently improving directions while contracting elsewhere, and $\sigma$ adapts the overall exploration scale. CMA-ES depends only on ranking and thus tolerates moderate noise, but its full-population evaluations and noisy fitness perturbations make it computationally expensive.

% ------------------------------------------------------------------------
\section{Proposed RACE-CMA Algorithm}
\subsection{Motivation}
Our objective $\mathcal{J}(\mathbf{T})$, which is a stochastic function estimated via Monte-Carlo simulations, is both expensive and noisy. Conventional optimizers exhibit complementary weaknesses: IPN converges rapidly on smooth landscapes but is local and noise-sensitive; SPSA is inexpensive per step but remains local and slow; CMA-ES is robust and global yet costly since each generation evaluates all $\lambda$ candidates in full. To address these challenges, we develop the Ranking-Aware, Constrained, and Efficient CMA-ES (RACE-CMA), which retains the CMA-ES backbone while significantly reducing simulation cost and sensitivity to noise through:
(i)~two-stage racing with common random numbers (CRN),
(ii)~uncertainty-weighted recombination, and
(iii)~feasible-by-construction constraint handling.
\subsection{Algorithmic Framework}

\paragraph{Two-Stage Racing with CRN}
At generation $t$, samples are drawn same as CMA-ES and mapped into a feasible ordered triple.  
\emph{Stage~1} performs a coarse evaluation of all $\lambda$ samples using a low-fidelity estimator $\tilde{\mathcal{J}}(\mathbf{T})$ (fewer Monte-Carlo trials), employing identical random seeds $s_1$ to maintain correlated noise and stabilize ranking:
\begin{equation}
\tilde{\mathcal{J}}\!\big(\mathbf{T}^{(j)}; s_1\big), \qquad j=1,\dots,\lambda.
\end{equation}
Let $\mathcal{K}_t$ denote the indices of the top $k=\rho\lambda$ candidates by $\tilde{\mathcal{J}}(\mathbf{T})$, where $\rho\in(0,1]$ is the promotion fraction.  
\emph{Stage~2} evaluates only these promoted candidates using the full simulator, with $r_j\!\ge\!1$ independent repetitions under CRN seeds $\{s_{2,\ell}\}$ to estimate their mean and variance:
\begin{equation}
\begin{aligned}
\hat{\mathcal{J}}_j = \tfrac{1}{r_j}\sum_{\ell=1}^{r_j} \mathcal{J}(\mathbf{T}^{(j)}; s_{2,\ell}), \, \\
\hat\sigma_j^2 = \tfrac{1}{r_j-1}\sum_{\ell=1}^{r_j} 
\big({\mathcal{J}}(\mathbf{T}^{(j)}; s_{2,\ell})-& \hat{\mathcal{J}}(\mathbf{T})_j\big)^2
\end{aligned}
\end{equation}
Using identical CRN seeds ensures all candidates face identical random perturbations, thereby reducing rank errors induced by noise.  
For non-promoted candidates, a worst-case variance
$\hat\sigma_j^2=\max_{i\in\mathcal{K}_t}\hat\sigma_i^2$
is assigned so that they preserve their Stage~1 order but are effectively excluded from elite updates. Let $C_f$ denote the cost of a full evaluation and $C_c$ that of a coarse Stage~1 evaluation, with $\tau=C_c/C_f\in(0,1)$.  
The per-generation simulation cost is then
\begin{equation}
\underbrace{\lambda\tau C_f}_{\text{Stage 1}} \;+\;
\underbrace{\rho\lambda\beta C_f}_{\text{Stage 2}}
=\lambda(\tau+\rho\beta)C_f,
\label{eq:cost}
\end{equation}
where $\beta\!\in\!(0,1]$ represents any early stopping or adaptive truncation in Stage~2.  
Thus, RACE-CMA reduces per-generation cost from $\mathcal{O}(\lambda C_f)$ in CMA-ES to $\mathcal{O}\!\big(\lambda(\tau+\rho\beta)C_f\big)$, which can be an order-of-magnitude saving when $\tau\ll1$ or $\rho\ll1$.

\paragraph{Uncertainty-Weighted Recombination}
To reduce the impact of noisy outliers, elite candidates are down-weighted during recombination.  
For the top $\mu$ promoted samples $x_{i:\lambda}$ with CMA-ES weights $w_i>0$, we apply inverse-variance weighting $\tilde w_i \propto \frac{w_i}{\epsilon+\hat\sigma_i^2},$ where $ 
\sum_{i=1}^{\mu}\tilde w_i=1,$ and $\epsilon\!\ll\!1$ avoids division by zero. Candidates with higher variance thus influence the mean and covariance updates less, improving stability under noise.  
The standard evolution-path and step-size updates remain unchanged.

\paragraph{Feasible Parameterization}
Threshold ordering is enforced via the differentiable mapping as in $T_1 = u_1, \,
T_2 = T_1+\delta+\mathrm{softplus}(u_2), \,
T_3 = T_2+\delta+\mathrm{softplus}(u_3),
$ with $\delta>0$ ensuring minimum spacing.  
This avoids post-hoc sorting, which distorts the sampling distribution, by guaranteeing feasibility directly.  An alternative is to optimize ordered RESI quantiles $q_1>q_2>q_3$ and map them via $T_i=F^{-1}_{\mathrm{RESI}}(q_i)$, eliminating constraint-correction bias.

\begin{algorithm}[t]
\caption{RACE-CMA Threshold Optimization}
\label{alg:RACECMA}
\begin{algorithmic}[1]
\Require Objective $f(\mathbf{T})$; estimator $\tilde f(\mathbf{T})$ with $\tau=C_c/C_f$; 
% feasible mapping $\mathcal{G}:\mathbb{R}^n\!\to\!\Omega$; population $\lambda$, elites $\mu$, weights $w_i>0$ ($\sum_i w_i=1$); promotion fract. $\rho\in(0,1]$; 
% learning rates $c_\sigma,d_\sigma,c_c,c_1,c_\mu$; CRN seeds $s_1,\{\bm{s}_{2,\ell}\}$.
\State \textbf{init} $m^{(0)}$, $\sigma^{(0)}>0$, $C^{(0)}\!=\!I$, $p_\sigma=0$, $p_c=0$; set $g\!\leftarrow\!0$.
\While{not stopped $|| \; g \; < \; G$}
  \State $C^{(g)} = AA^\top$ (Cholesky/Eigen Decomposition).
  \For{$j=1{:}\lambda$} \Comment{offspring}
     \State $z_j\sim\mathcal{N}(0,I)$;\quad $u^{(j)} \leftarrow m^{(g)} + \sigma^{(g)}A z_j$.
     \State $\mathbf{T}^{(j)} \leftarrow \mathcal{G}(u^{(j)})$;\quad $\tilde f_j \leftarrow \tilde f(\mathbf{T}^{(j)};s_1)$.
  \EndFor
  \State Rank $\tilde f_j$; keep top $k=\lfloor \rho\lambda\rfloor$ indices $\mathcal{K}$.
  \For{$j\in\mathcal{K}$} \Comment{Stage-2}
     \State Evaluate $r_j\!\ge\!1$ times: $\hat f_j, \hat\sigma_j^2 \leftarrow f(\mathbf{T}^{(j)};s_{2,\ell})$.
  \EndFor
  \For{$j\notin\mathcal{K}$}  \State set $\hat f_j \leftarrow \tilde f_j+\varepsilon$, $\hat\sigma_j^2 \leftarrow \max_{i\in\mathcal{K}}\hat\sigma_i^2$.
  \EndFor
  \State Rank all $\hat f_j$; denote elites $x_{i:\lambda}$, $y_i = \frac{(x_{i:\lambda}-m^{(g)})}{\sigma^{(g)}}$.
  \State Uncertainty-weighted and normalised $\tilde w_i \propto \frac{w_i}{(\epsilon+\hat\sigma_{i:\lambda}^2)}$.
  \State $m^{(g+1)} \leftarrow \sum_{i=1}^\mu \tilde w_i x_{i:\lambda}$;\quad $y=(m^{(g+1)}-m^{(g)})/\sigma^{(g)}$.
  \State $p_\sigma \leftarrow$ Eq.~\eqref{eq:pq};\; $\sigma^{(g+1)}\leftarrow$ Eq.~\eqref{eq:sig};\; $p_c \leftarrow$ Eq.~\eqref{eq:pc}
  \State $C^{(g+1)} \leftarrow$ Eq.~\eqref{eq:up} with $\tilde w_i$; \; $g\!\leftarrow g+1$.
\EndWhile
\State \Return best $\mathbf{T}$ seen.
\end{algorithmic}
\end{algorithm}

\paragraph{Structured Sampling and Covariance Simplification}
To further reduce computational burden, we employ orthogonal and mirrored sampling so that offspring directions are symmetric.  
This structured design improves coverage of the search space and cancels odd-order noise, enabling smaller population sizes $\lambda$ without performance degradation.  
During early exploratory phases, the covariance matrix is restricted to a diagonal form while the global step size $\sigma$ remains large; full covariance adaptation is reactivated once convergence narrows the search region.  
These modifications reduce unnecessary correlation learning and lower the evaluation budget without impairing convergence quality.

\subsection{Discussion and Computational Analysis}

RACE-CMA achieves a substantial reduction in total simulation cost while maintaining strong global exploration ability. The per-generation cost ratio satisfies
\begin{equation}
\frac{C_{\text{RACE-CMA}}}{C_{\text{CMA-ES}}} = \tau + \rho\beta \ll 1,
\end{equation}
since standard CMA-ES requires $\lambda$ full evaluations per generation, whereas RACE-CMA reduces this by combining cheap coarse evaluations ($\tau=C_c/C_f$) with selective promotions controlled by $\rho$. CRN enforces consistent candidate ranking under noise, inverse-variance weighting mitigates outliers, and feasible reparameterization guarantees thresholds order without bias. 
% With typical settings $(\lambda,\mu,\rho,\tau,\beta)=(12,6,0.5,0.2,0.8)$, we achieved over a 50\% reduction in equivalent simulation cost while preserving accuracy and robustness.

\subsection{Extension to General Configuration Tuning}
In this work, we focus on threshold optimization as a representative case study. However, RACE-CMA readily generalizes to broader ISAC configuration tuning. Tunable parameters (e.g., beam-sweep cadence or RS density) can be modeled as continuous or ordered variables within network-defined bounds. Feasible-by-construction parameterization extends to vector-valued configurations using standard mappings (e.g., logistic/softplus for bounded variables, circular encoding for periodic ones, and relaxed embeddings for discrete choices). Under this formulation, RACE-CMA samples feasible configurations, evaluates their stochastic KPIs, and adapts its covariance to capture dominant sensitivities. This enables scalable multidimensional tuning while preserving the efficiency and robustness of the proposed approach.

\section{Simulation Results and Discussion}

\label{sec:results}

This section presents the performance evaluation of the proposed Configuration Tuning framework and RACE-CMA algorithm. Two categories of results are discussed: (i) algorithm-level evaluation, and (ii) system-level evaluation. The bistatic ISAC setup includes one BS and one sensing UE, operating with the parameters noted in Table ~\ref{tab:params} ($N_{\text{BS}}$ and $N_{\textbf{UE}}$ denote the number of antennas on BS and UE while $M_{tg}$ is the number of targets). All methods use CRN for fairness and RACE-CMA is configured with $\lambda=12$, $\mu=6$, $\rho=0.5$, $\tau=0.2$, and $\beta=0.8$. These parameters (except $\tau$) are selected based on standard CMA-ES empirical tuning, same as in \cite{Hanz2023CMA}, providing a stable trade-off between exploration, noise robustness, and computational cost.
The evaluation compares optimization performance (here detection reliability) under varying transmit-power budgets, focusing on convergence, robustness, and computational efficiency\footnote{In this study, a Monte Carlo simulation is used to capture key aspects of bistatic ISAC. While it enables comparisons, it doesn't reflect all real-world impairments. Further validation with higher-fidelity models is future work.}.

\renewcommand{\arraystretch}{1}
\rowcolors{1}{gray!15}{white}
\begin{table}[]
\centering
\caption{Simulation Parameters}
\begin{tabular}{
  >{\centering\arraybackslash}m{1.6cm} 
  >{\centering\arraybackslash}m{1.8cm}| 
  >{\centering\arraybackslash}m{1.8cm} 
  >{\centering\arraybackslash}m{1.8cm}}
\toprule
$[N_{\text{BS}}, N_{\textbf{UE}}]$       & [32, 16]            & BS Beams & 20 
\\ Antenna spacing.           &  $\frac{\lambda_c}{2}$       &  Sweep Range        & $[\frac{\pi}{4}, \frac{3\pi}{4}]$ 
\\ $[f_c ,W_c]$       & [24G, 15k] Hz        & BS Power Rng. & $[10, 30] \text{ dBm}$ 
\\ Noise Figure         & 6 dB          & $[T_{\text{sym}}, N_{\text{sym}}]$            & $[100 \mu \text{s}, 100]$ 
\\ $[M_{tg}, v_{tg}]$ & $[1, 3 \text{ m/s}]$  & 
$[N_{\text{sub}}, W_{\text{sub}}]$ & $[4,10 W_c]$ 
\\ $T_\text{S}$ & 10 sec & $[N_{\text{del}}, N_{\text{dop}}]$ & [10, 10] \\
\bottomrule
\end{tabular}
\label{tab:params}
\end{table}
\begin{figure*}[!t]
  \centering
  \subfloat[ Detection reliability]{%
    \includegraphics[width=0.32\linewidth]{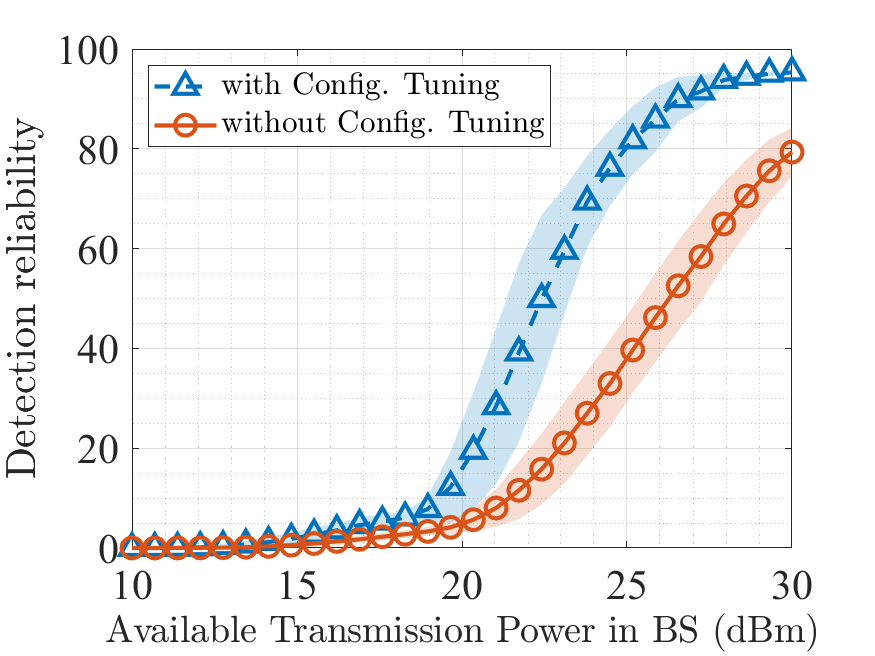}
    \label{fig:pd}}
  \hfill
  \subfloat[Latency]{%
    \includegraphics[width=0.32\linewidth]{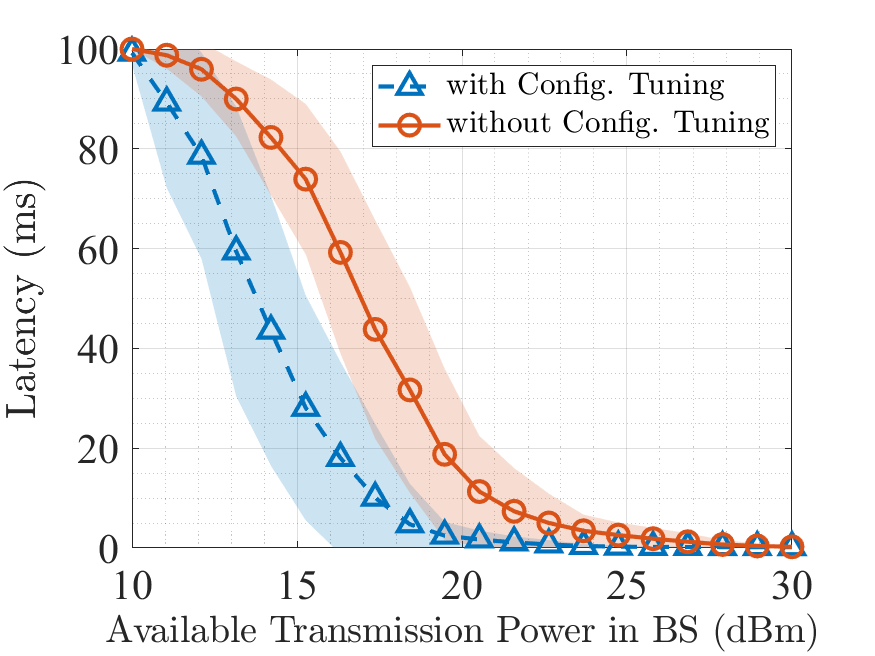}
    \label{fig:laten}}
  \hfill
  \subfloat[S\&C power portion from Budget]{%
    \includegraphics[width=0.32\linewidth]{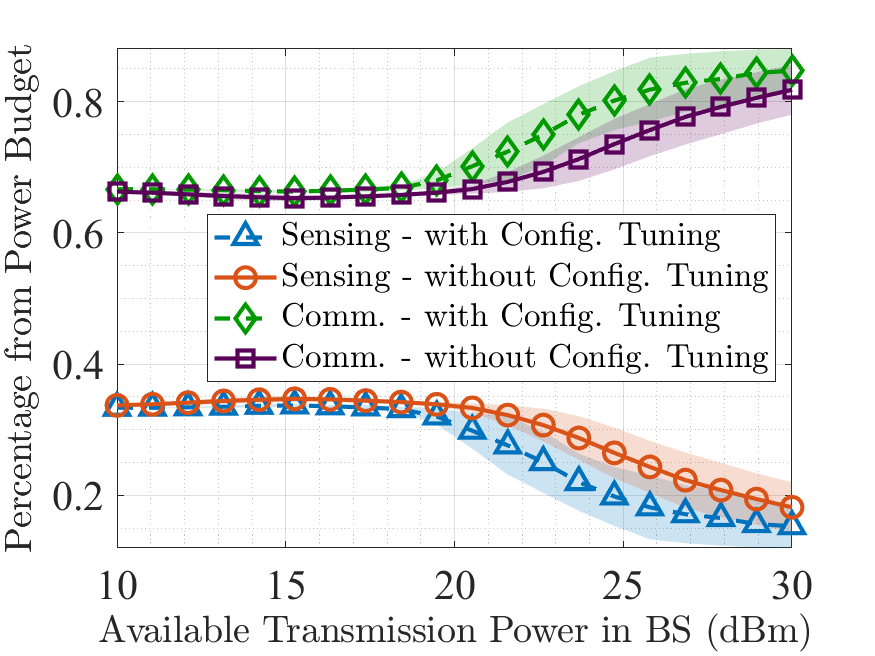}
    \label{fig:ratio}}
  \caption{Overall gains (mean $\pm$ 95\% CI) from configuration tuning under randomized UE positions and initial configurations: (a) improved  Detection reliability, (b) reduced sensing-feedback latency, and (c) adaptive allocation of the BS power budget toward sensing without degrading communication throughput.}
  \label{fig:system}
\end{figure*}

\begin{table}
\centering
\caption{Comparison across $100$ runs (mean $\pm$ 95\% CI)}
    % \resizebox{\columnwidth}{!}{% Make table fit column width

\begin{tabular}{lccccc}
\toprule
Method & impr. $\Delta J$ & Cost ($N_{\text{eq}}$) & Efficiency ($\frac{\Delta J}{N_{\text{eq}}} $ ) \\
\midrule
IPN       & 10\% $\pm$ 1\% & 60 $\pm$ 2 & 0.16 \\
SPSA      & 8\% $\pm$ 0.5\% &  65 $\pm$ 5 & 0.1 \\
CMA-ES    & 24\% $\pm$ 3\% & 96 $\pm$ 6 & 0.25 \\
\textbf{RACE-CMA} 
          & \textbf{35\% $\pm$ 2\%} 
          & \textbf{72 $\pm$ 4} 
          & \textbf{0.48} \\
\bottomrule
\end{tabular}
% }
\label{tab:comparison}
\end{table}
% --------------------------------------------------------------
\subsection{Algorithm Evaluation}
We first assess the performance of the proposed RACE-CMA against IPN, SPSA, and CMA-ES under varying transmit power budgets.  
The analysis considers convergence dynamics against power variations, and computational efficiency.

\begin{figure}
\centering
\subfloat[Available Tx Power = $20$ dBm]{%
\includegraphics[width=0.49\linewidth]{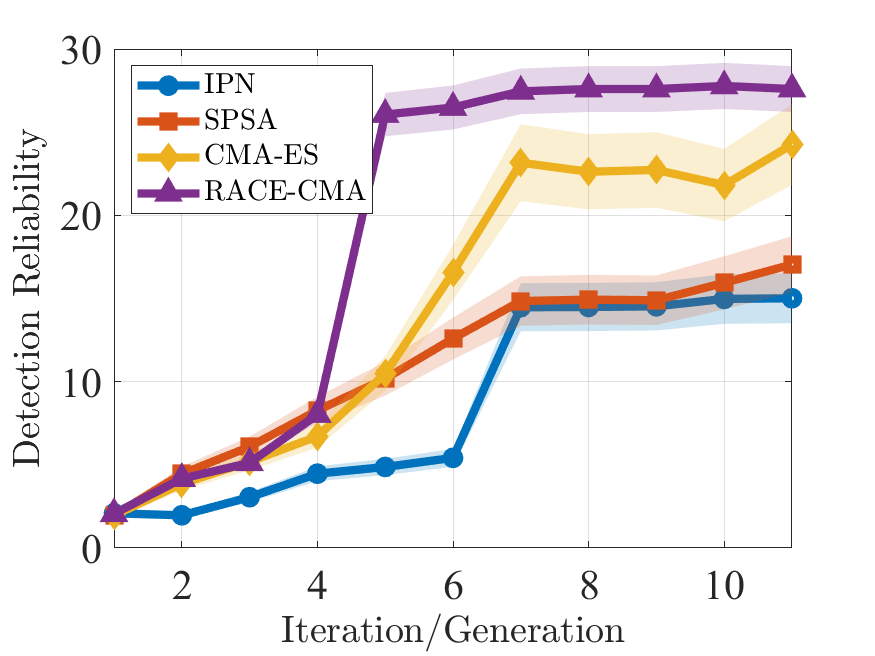}}
\hfill
\subfloat[Available Tx Power = $24.7$ dBm]{%
\includegraphics[width=0.49\linewidth]{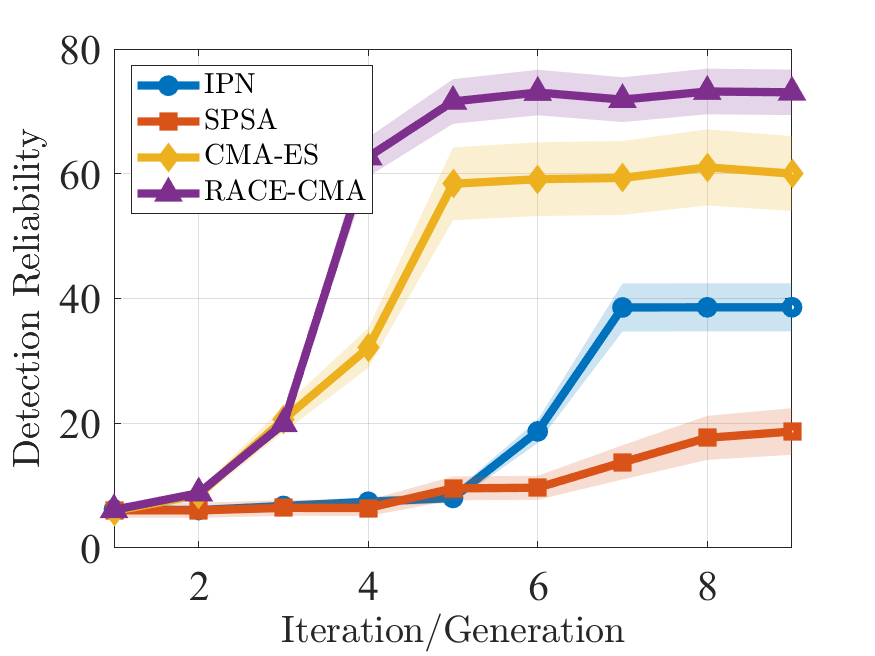}}
\caption{Convergence under different power budgets (mean $\pm$ 95\% CI)}
\label{fig:convergence}
\end{figure}

\paragraph{Convergence Behavior}
Fig.~\ref{fig:convergence}(a)–(b) show the evolution of $J_{\mathrm{det}}(\mathbf{T})$ over ten generations for two power budgets ($20$~dBm and $24.7$~dBm). At low power, all methods start weak, but RACE-CMA reaches its steady value within two to three generations, whereas CMA-ES requires nearly twice as long and IPN/SPSA remain suboptimal. At higher power, RACE-CMA attains near-perfect reliability by generation four, while CMA-ES remains roughly $50\%$ lower. These results demonstrate RACE-CMA’s faster and more reliable convergence across power levels.

\paragraph{Computational Efficiency}
Table~\ref{tab:comparison} summarizes quantitative comparisons averaged over 100 runs. RACE-CMA achieves the largest relative improvement ($35\%\!\pm\!2\%$) while requiring fewer equivalent simulations than CMA-ES. Its efficiency, measured as $\Delta J/N_{\mathrm{eq}}$, reaches $0.48$, nearly double that of CMA-ES and more than three times that of SPSA and IPN. Overall, RACE-CMA delivers the best trade-off between accuracy, cost, and efficiency among all methods.

% --------------------------------------------------------------
\subsection{System Evaluation}
\label{sec:syseval}

To assess end-to-end ISAC gains, the tuned parameters are applied in the full closed-loop sensing–communication system. Fig.~\ref{fig:system}(a)–(b) show that configuration tuning improves Detection reliability and significantly reduces latency by up to 60\% in low-power regimes (around 15 dBm). Threshold learning also raises the lower bound of both curves, yielding higher confidence bands compared with fixed-threshold operation. Fig.~\ref{fig:system}(c) shows the resulting resource allocation behavior: as the BS power budget increases, configuration tuning autonomously assigns a larger share of power to communications without harming sensing performance. Overall, configuration tuning improves sensing accuracy and responsiveness while enabling more efficient joint resource use in ISAC systems.

% \begin{figure}[h]
% \centering
% \subfloat[Detection and missed detection vs distance]{%
% \includegraphics[width=\linewidth]{Figs/p_det_from_Budget_randomUE_randomT.eps}}
% \\
% \subfloat[Cumulative latency distribution]{%
% \includegraphics[width=\linewidth]{Figs/Latency_from_Budget_randomUE_randomT.eps}}
% \caption{Overall performance gains from threshold optimization under randomized UE positions and initial thresholds.}
% \label{fig:system}
% \end{figure}
\textbf{Note.}
The results focus on sensing KPIs and resource allocation trends. While communication is not adversely affected (Fig.~\ref{fig:system}(c)), detailed metrics (e.g., SINR) are future work.

% {\color{red}{\paragraph*{Scope and limitations}
% The system-level evaluation focuses primarily on sensing-side performance (reliability and latency) and on resource allocation trends. While Fig.~3(c) illustrates that communication resources are not adversely impacted in the considered setup, a detailed communication-centric performance analysis (e.g., throughput or SINR distributions) is beyond the scope of this work. Similarly, the sensing model is intentionally simplified (bistatic, RESI-based, limited number of targets) to isolate the configuration-tuning problem. Extending the evaluation to richer communication metrics and more complex sensing scenarios is an important direction for future work.}}

\section{Conclusion}

This work presented configuration tuning as a scalable and standards-compliant mechanism for adaptive closed-loop ISAC operation. By casting sensing configuration adaptation as a stochastic constrained optimization problem, we proposed RACE-CMA, a noise-robust and simulation-efficient evolutionary optimizer. Simulation results show that RACE-CMA improves sensing reliability by up to 35\% while reducing equivalent simulation cost by over 50\%, alongside significant latency reduction and more efficient sensing–communication resource allocation. These results demonstrate the potential of configuration tuning for improving cost-efficiency and reliability in closed-loop ISAC systems under the considered modeling assumptions.
% establish configuration tuning as a practical enabler for cost-efficient and reliable ISAC adaptation in future radio access networks.

\bibliographystyle{IEEEtran}  % Use IEEE bibliography style
\bibliography{References}  % Name of your .bib file (without .bib extension)
\end{document}